# Adaptive Inverse Mapping: A Model-free Semi-supervised Learning Approach towards Robust Imaging through Dynamic Scattering Media

Xiaowen Hu[1], Jian Zhao[1,2*], Jose Enrique Antonio-Lopez[1], Stefan Gausmann[1], Rodrigo Amezcua Correa[1], and Axel Schülzgen[1]

**Abstract**

Imaging through scattering media is a useful and yet demanding task since it involves solving for an inverse mapping from speckle images to object images. It becomes even more challenging when the scattering medium undergoes dynamic changes. Various approaches have been proposed in recent years. However, to date, none is able to preserve high image quality without either assuming a finite number of sources for dynamic changes, assuming a thin scattering medium, or requiring the access to both ends of the medium. In this paper, we propose an adaptive inverse mapping (AIP) method which is flexible regarding any dynamic change and only requires output speckle images after initialization. We show that the inverse mapping can be corrected through unsupervised learning if the output speckle images are followed closely. We test the AIP method on two numerical simulations, namely, a dynamic scattering system formulated as an evolving transmission matrix and a telescope with a changing random phase mask at a defocus plane. Then we experimentally apply the AIP method on a dynamic fiber-optic imaging system. Increased robustness in imaging is observed in all three cases. With the excellent performance, we see the great potential of the AIP method in imaging through dynamic scattering media.

[1] CREOL, The College of Optics and Photonics, University of Central Florida, Orlando, FL 32816, USA
[2] Department of Electrical and Computer Engineering, Boston University, Boston, MA 02215, USA
*jianzhao@knights.ucf.edu

Optical imaging through scattering media[1–3] is indispensable to many applications, ranging from underwater imaging[4] and biological tissue imaging[5] to imaging through atmosphere[6] and non-line-of-sight imaging[7,8]. Unfortunately, light from the object undergoes multiple scattering and forms a noise-like speckle image at the detector[9]. An inverse problem must be solved in order to retrieve the object image. The problem becomes much harder when the dynamic nature of scattering media has to be incorporated, for example, in living tissues[10–12]. Time-varying scattering properties rapidly scramble the optical information and result in decorrelations of the speckle images. Many approaches have been proposed over the past decades to address this issue. Yet none of them work satisfactorily. In the methods utilizing phase conjugation[13,14], wavefront shaping[15], or transmission matrix (TM)[16–18], the dynamic change is compensated by fast spatial light modulators (SLMs)[19] or deformable mirror devices (DMDs)[20,21] in feedback control techniques. Although these methods are flexible to any dynamic change, they require the access to both ends of the scattering medium, which is often unattainable in real-world applications. Deep learning methods[22–25] establish a robust inverse mapping by training a convolutional neural network (CNN) with numerous pairs of speckle and object images collected under different conditions[26–30]. After training, the CNN can reconstruct object images from speckle images. Nevertheless, since the CNN is fixed at the time of the test, any dynamic change that has not been included during training can significantly degrade the reconstructed images. Memory effects for speckle correlation[31–33] enable single-shot imaging, overcoming the disadvantages of the above methods. However, the memory effect method is limited to thin films. As the scattering media get thicker, the speckle correlation drops rapidly and memory effects become neglectable.

Here, we present an adaptive inverse mapping (AIP) method that is flexible regarding any dynamic change while it requires only the speckle images for the object image recovery. We improve the deep learning approach by making the inverse mapping adaptive to the dynamic scattering medium. We show that although the reconstructed images are scrambled in the dynamic scattering medium, their relationship to the object images is preserved. By utilizing recently developed unpaired image-to-image translation[34], the AIP method is able to correct the inverse mapping and stabilize the imaging performance. As proofs of concept, we test the AIP method on two numerical simulations: a dynamic scattering medium formulated as an evolving TM, and a telescope with a dynamic random phase mask at a defocused plane. We then experimentally apply the AIP method to a glass-air Anderson localization optical fiber (GALOF) based imaging system with a changing imaging depth and small displacement of optical components. By closely monitoring the output speckle images, the AIP method preserves a high quality of reconstructed images in all three cases. With the universality shown, we see the great potential of the AIP method on increasing robustness in imaging through a wide range of dynamic scattering media.

**Results**

**Principle** We illustrate the AIP method in Fig. 1. At any 'snapshot' of the dynamic scattering medium, e.g., state *i-1*, the scattering medium can be viewed as a forward mapping that takes object images and outputs speckle images. An inverse mapping is then applied to reverse this process to reconstruct object images. At the next state *i*, dynamic changes in the scattering medium perturb the forward mapping so that distorted images will be generated if the same inverse mapping is used. The AIP method corrects the inverse mapping by monitoring the output speckle images from the scattering medium (Fig. 1 (a)). More specifically, we initialize an inverse mapping of the scattering medium at state 0 by training a $CNN_0$ on *m* pairs of speckles and ground truth object images (Fig. 1 (b)). At any subsequent state *i*, *m* output speckle images are passed through the inverse mapping of the previous state $CNN_{i-1}$, generating *m* distorted reconstructed images (Fig. 1 (c)). Among those *m* distorted images, *n* images are used, together with *n* reserved object images, to train an image restoration cycle-consistent adversarial network *i* (Restore-CycleGAN$_i$ (Fig. 1 (c) blue box and arrow). Note that not only are these two sets of images unpaired, but the true objects of the distorted images do not need to be included in the object images. After learning a translation between distorted images and object images, the Restore-CycleGAN$_i$ takes all the *m* distorted images and generates *m* object images. In this way, we have *m* pairs of speckle and object images. Finally, a CNN$_i$ learns a mapping from the *m* speckle images to the *m* object images (Fig. 1 (c) green boxes and arrow). As a result, the inverse mapping is adapted to the new state *i*. The AIP method is semi-supervised in the sense that paired images are only required at the initialization. In later states, all it needs is output speckle images.

Fig. 1 (d) shows the detailed flowchart of a Restore-CycleGAN. Similar to the original CycleGAN[34], two generator networks $G_{obj}$ and $G_{dis}$ try to learn a mapping from the distorted reconstructions to the objects and vice versa. Two discriminators $D_{obj}$ and $D_{dis}$ try to distinguish between the real images in the target domain and the fake images from the generators. The generators and discriminators are optimized in an adversarial game through the least square adversarial loss $\mathcal{L}_{\text{LSGAN}}$. Further, the generators are optimized through two more losses: the cycle-consistent loss $\mathcal{L}_{\text{cycle}}$ and the identity mapping loss $\mathcal{L}_{\text{identity}}$. $\mathcal{L}_{\text{cycle}}$ requires that an image should be unaltered if it goes through a full cycle. $\mathcal{L}_{\text{identity}}$ imposes an identical output if the input is an image from the target domain. We use UNets[35] as the generators in place of ResNets[36] in the CycleGAN[34], inspired by the observation that UNets with skip-connections have weaker priors than ResNets[37]. This enhances the performance of the Restore-CycleGAN on image restoration (see Methods for a comparison to CycleGAN). PatchGANs[38] are used as the discriminators. The details of the architecture and the training process of Restore-CycleGANs can be found in Methods.

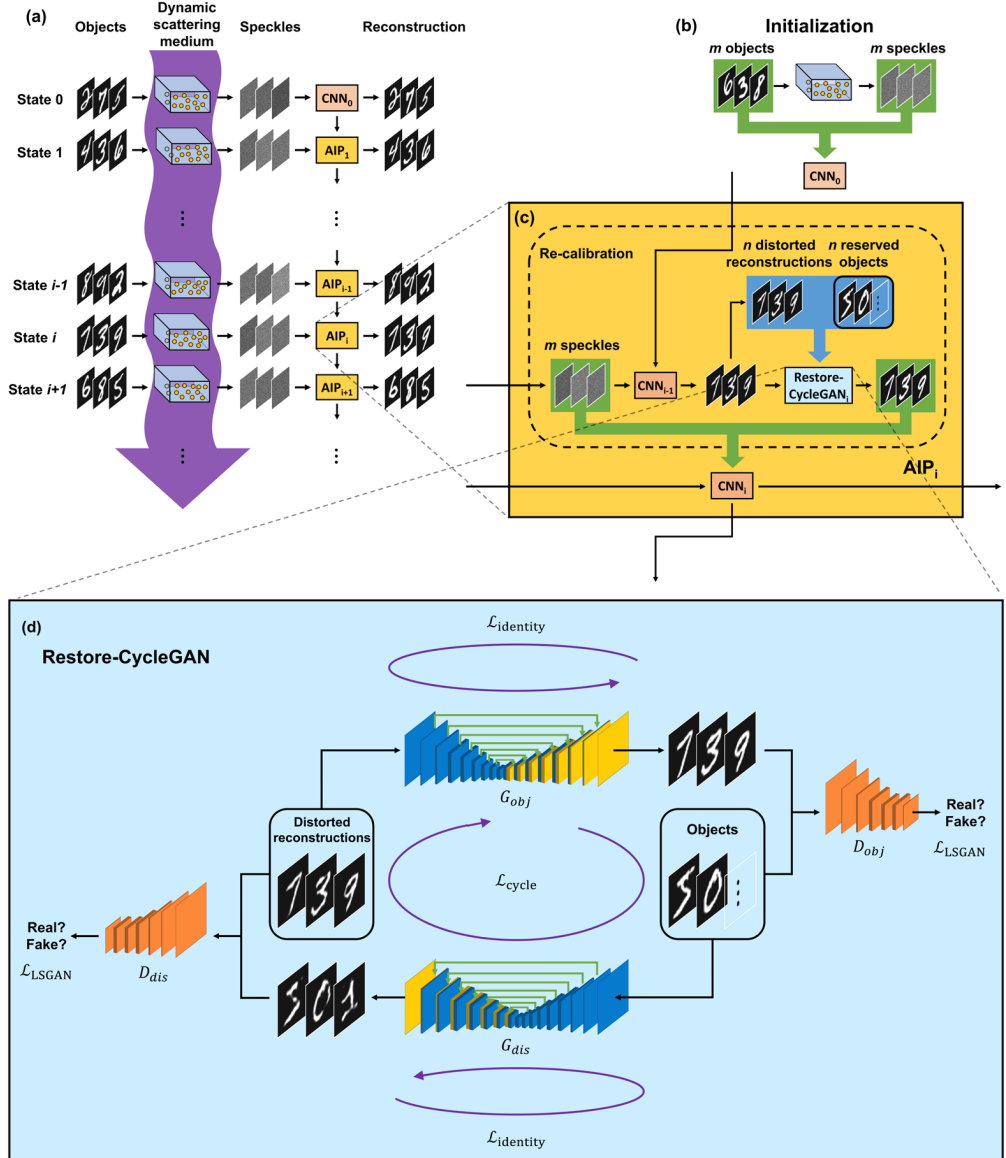

Fig. 1 (a) The diagram of applying the AIP method on imaging through dynamic scattering medium. The AIP method constantly corrects the inverse mapping by monitoring output speckle images from 'snapshot' states of the scattering medium. (b) Initialization of the AIP method. At state 0, $m$ object images are passed through the scattering medium, generating $m$ speckle images. A $CNN_0$ is trained on these $m$ pairs of images (green boxes) to establish an initial inverse mapping. (c) The detailed workflow of the $AIP_i$. In the re-calibration stage (dashed box), $m$ output speckles pass through the inverse mapping $CNN_{i-1}$ of the previous state $i-1$. Because of the dynamic changes in the scattering medium, $CNN_{i-1}$ generates $m$ distorted reconstructions. $n$ images are randomly chosen among these $m$ images, together with n reserved object images, to form the training set of Restore-$CycleGAN_i$ (blue box). After learning a transition between those unpaired images, the Restore-$CycleGAN_i$ takes all the $m$ distorted images and generates $m$ clear images. Therefore, we have $m$ pairs of speckle images and object images (green boxes). Finally, a $CNN_i$ is trained on these $m$ pairs of images to re-establish the inverse mapping at the state $i$. Blue box: unsupervised learning. Green boxes: supervised learning. (d) The flowchart of the Restore-CycleGAN in (c). It consists of two generator-discriminator pairs: the object image generator $G_{obj}$ and the discriminator $D_{obj}$, and the distorted image generator $G_{dis}$ and the discriminator $D_{dis}$. The least square adversarial loss $\mathcal{L}_{\text{LSGAN}}$ is optimized in a min-max game, in which $G_{obj}$ tries to fool $D_{obj}$ by generating object images from distorted images, whereas $D_{obj}$ distinguishes between the real objects and the fake objects generated by $G_{obj}$. Similarly, there is a $\mathcal{L}_{\text{LSGAN}}$ for the reversed direction. The cycle-consistent loss $\mathcal{L}_{\text{cycle}}$ enforces an identical output if an image passes through a full translation cycle. The identity mapping loss $\mathcal{L}_{\text{identity}}$ regularizes the generators to have an identity mapping if the input is a real image from the target domain.

**Dynamic scattering imaging system as an evolving TM** We test the AIP method on a general case of dynamic scattering imaging systems, where the system is formulated as a complex-valued TM relating the input image to the output image (see Methods for details). We construct a TM by drawing its elements from a complex normal distribution[39–42] with a zero mean and a variance 1, i.e. $CN(0,1)$. Dynamic changes are introduced by gradually replacing the elements in the TM with new elements from the same complex normal distribution (Fig. 2(a)). The imaging objects are Modified National Institute of Standards and Technology (MNIST) handwritten digits[43] resized to 256x256. Starting from an initial inverse mapping $CNN_0$, the AIP method is applied with $m$=5000 and $n$=1000 every time when the percentage of the substituted elements $p$ in the TM is increased by $\Delta p$=12.5%. The performance of image reconstruction at all states is evaluated on a separate set of 500 test images. The results are shown Fig. 2(b-d). As the $p$ increases, the reconstructed images by $CNN_0$ become more and more unrecognizable (Fig. 2 (b)). In comparison, the AIP method stabilizes image reconstruction by improving the inverse mapping based on their precedents. The improvement is quantified in Fig. 2 (c), where we plot the averages and standard deviations of the mean absolute errors (MAEs) between the reconstructions and the objects at different states. For every $AIP_{i-1}$, the MAE increases when the dynamic scattering system transforms to a later state $i$. The $AIP_i$ then corrects the inverse mapping and lowers the MAE. Fig. 2 (d) shows the output speckle decorrelations as a function of $p$. The amount of speckle decorrelation is evaluated through the Pearson correlation coefficient (PCC). When the output speckles at $p$=50% are already decorrelated to the speckles at the first state, e.g., PCC<1/e, good image reconstruction is still preserved. We attribute this to the fact that the speckles remain highly correlated to the neighboring state if the system is traced closely. This is further confirmed through a comparison with the results by the AIP method with an increased $\Delta p$=25% (Methods).

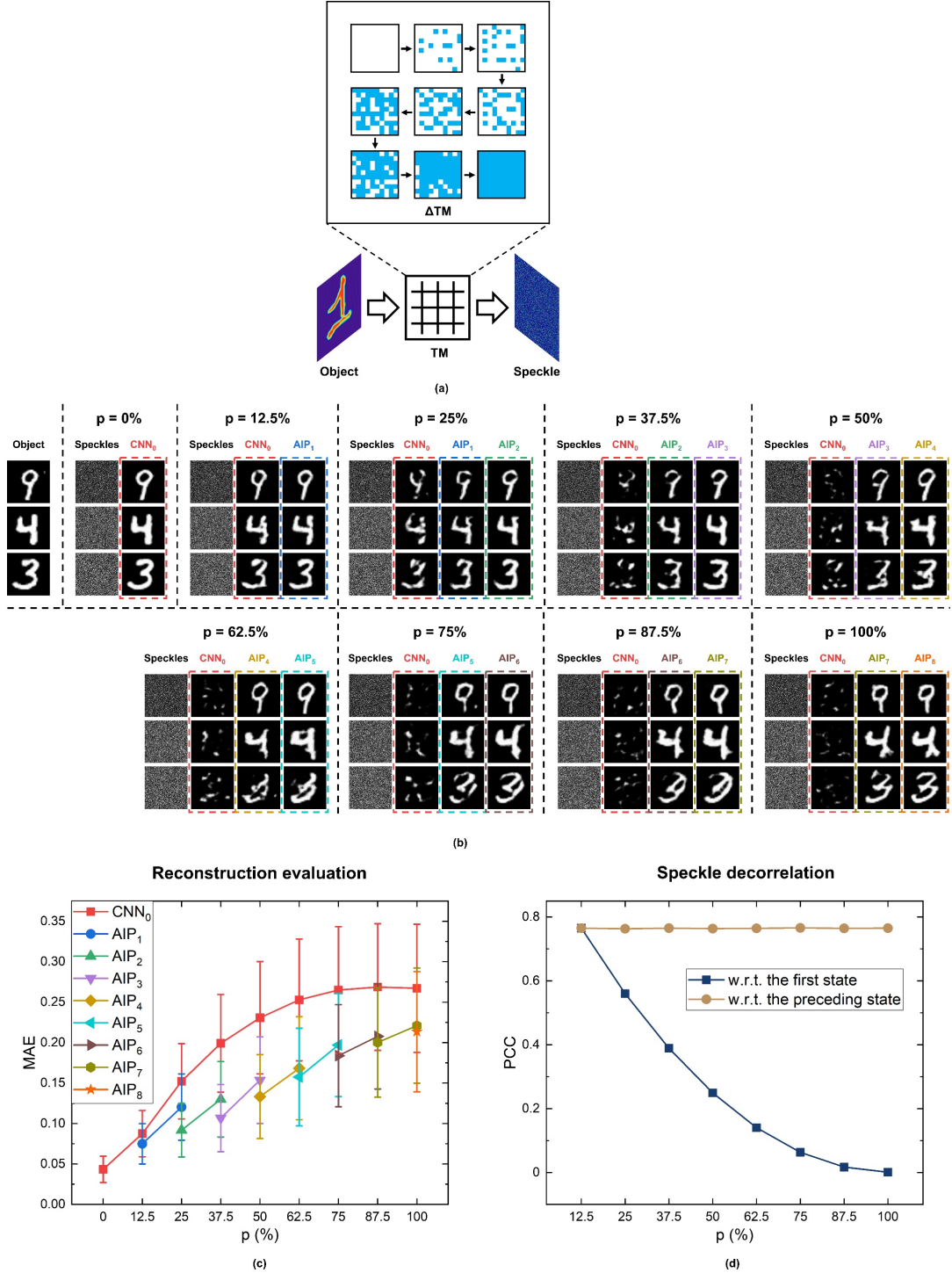

Fig. 2 (a) Schematic of a dynamic scattering system formulated as an evolving TM. The TM transforms objects into speckles. Enlarged figure: The dynamic changes in the TM. The elements in the original TM (white) are gradually replaced by new elements (blue). (b) Image reconstructions results obtained by the AIP method to the dynamic scattering system shown in (a). Object: The input object images (top left column); Speckle: the output speckle images when the percentage of the substituted elements $p$ in the TM is increased from 0% to 100% with a step of 12.5%; $CNN_0$ and $AIP_i$: the reconstructed images from $CNN_0$ and the $i^{th}$ AIP. The dashed bounding box groups the reconstructions from a particular AIP or $CNN_0$. (c) The averages and standard deviations of the MAE of the test reconstructions from the $AIP_i$ and $CNN_0$. The colors of the symbols correspond to the colors of the bounding boxes in (b). (d) The PCCs of the speckle images with respect to (w.r.t.) the speckle images from the first state $p$=0% (dark blue line), and w.r.t. the speckle images from the preceding state (brown line).

**Dynamic telescopic imaging system** Next, we numerically simulate the use of the AIP method on a dynamic telescopic imaging system, where a changing random phase mask is located at a defocused plane[25,26,30,44,45] (Fig. 3). The focal lengths $f_1$ and $f_2$ of the two lenses are chosen to be 250 mm and 150 mm, respectively. The object has a size of 10.24 mm x 10.24 mm. A random phase mask is placed z = 15 mm in front of the first lens L1. The transmittance of the phase mask $t(x,y)$ is formulated as Eqs. (1-2) given in Fig. 3. $\Delta n = 0.52$ is the refractive index difference between the phase mask and air. $\lambda = 632.8\text{nm}$ is the wavelength. $D(x,y)$ is a random height field. $W(x,y)$ is a set of random height values drawn from the normal distribution $N(\mu, \sigma_0)$ at discrete sample location $(x,y)$, and $K(\sigma)$ is a zero-mean Gaussian smoothing kernel with a full width half maximum (FWHM) of $\sigma$. Moreover, the elements in the matrix $W(x,y)$ are gradually replaced by values from the same normal distribution (enlarged figure in Fig. 3). Thus, the phase mask is changing towards a different phase mask. $\mu$, $\sigma_0$, $\sigma$ are chosen to be 16 μm, 5 μm and 4 μm, respectively. Imaging through the system is simulated using Fourier optics.

The imaging objects are extended MNIST (EMNIST) handwritten letters[46]. We initialize an inverse mapping $CNN_0$ of the system with the original phase mask. The AIP method is then adopted to stabilize the imaging reconstruction every time when the percentage of the substituted elements *p* in the $W(x,y)$ is increased by *Δp*=10%. We choose *m*=5000, and *n*=1000 in the AIP method. A separate set containing 500 images is used to test the performance of image reconstruction at all states. The results are shown in Fig. 4. As *p* increases, the quality of images reconstructed by $CNN_0$ degrades (Fig. 4 (a)). In comparison, the AIP method stabilizes the image reconstruction (the last column at each state in Fig. 4 (a)). Good visual quality is still maintained when the original phase mask has been completely replaced by a new phase mask (*p*=100%). In Fig. 4 (b), we plot the averages and standard deviations of the MAE between the reconstructions and the objects at different states. Similar trends as in Fig. 2 (c) can be observed, where the AIP method corrects the inverse mapping of the preceding state. Fig. 4 (c) shows the averages and standard deviations of the PCC scores between the output speckles. While the speckles become more and more decorrelated from the speckles at the first state (dark blue line), they remain highly correlated with the speckles from the preceding states (brown line). This indicates the necessity of tracing the state of the system closely. We compare to the results when the AIP method is applied directly at the final state. Degradation on the reconstructions can be seen from both the additional columns in Fig. 4 (a) and the dark gray star in Fig. 4 (b) at *p* = 100%.

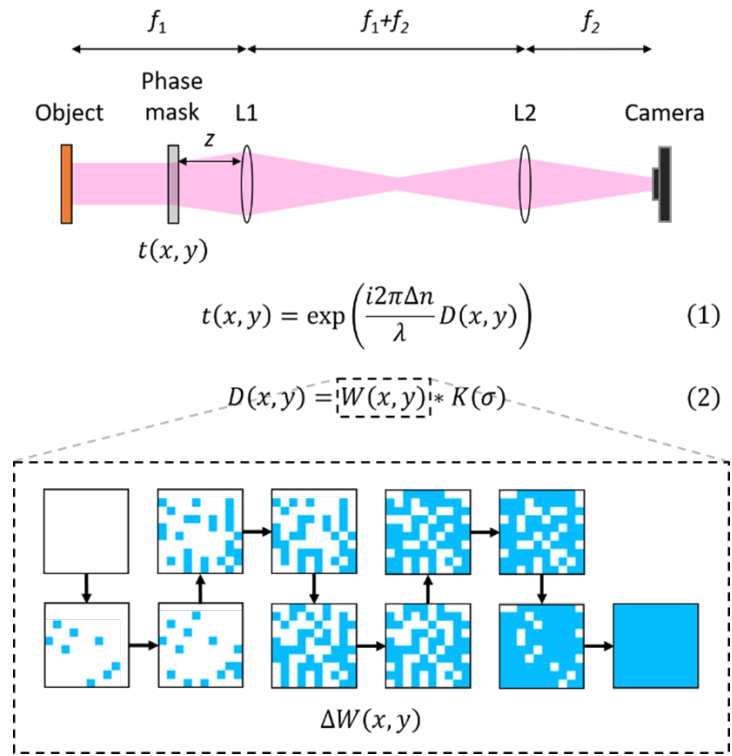

$$t(x,y) = \exp\left(\frac{i2\pi\Delta n}{\lambda}D(x,y)\right) \quad (1)$$

$$D(x,y) = W(x,y) * K(\sigma) \quad (2)$$

Fig. 3 Schematic of the simulated dynamic telescopic imaging system. A changing random phase mask is placed at a defocused plane. L1, L2: lenses. Eq. (1-2): formulas of the transmittance of the random phase mask $t(x,y)$. $\Delta n$ : the refractive index difference between the phase mask and air. $\lambda$ : wavelength. $D(x,y)$ : a random height field. $W(x,y)$ : a set of random height values drawn from the normal distribution $N(\mu,\sigma_0)$ at discrete sample locations $(x,y)$. $K(\sigma)$ : a zero-mean Gaussian smoothing kernel with FWHM value of $\sigma$. Enlarged: the elements in $W(x,y)$ (white) are gradually replaced by new elements (blue).

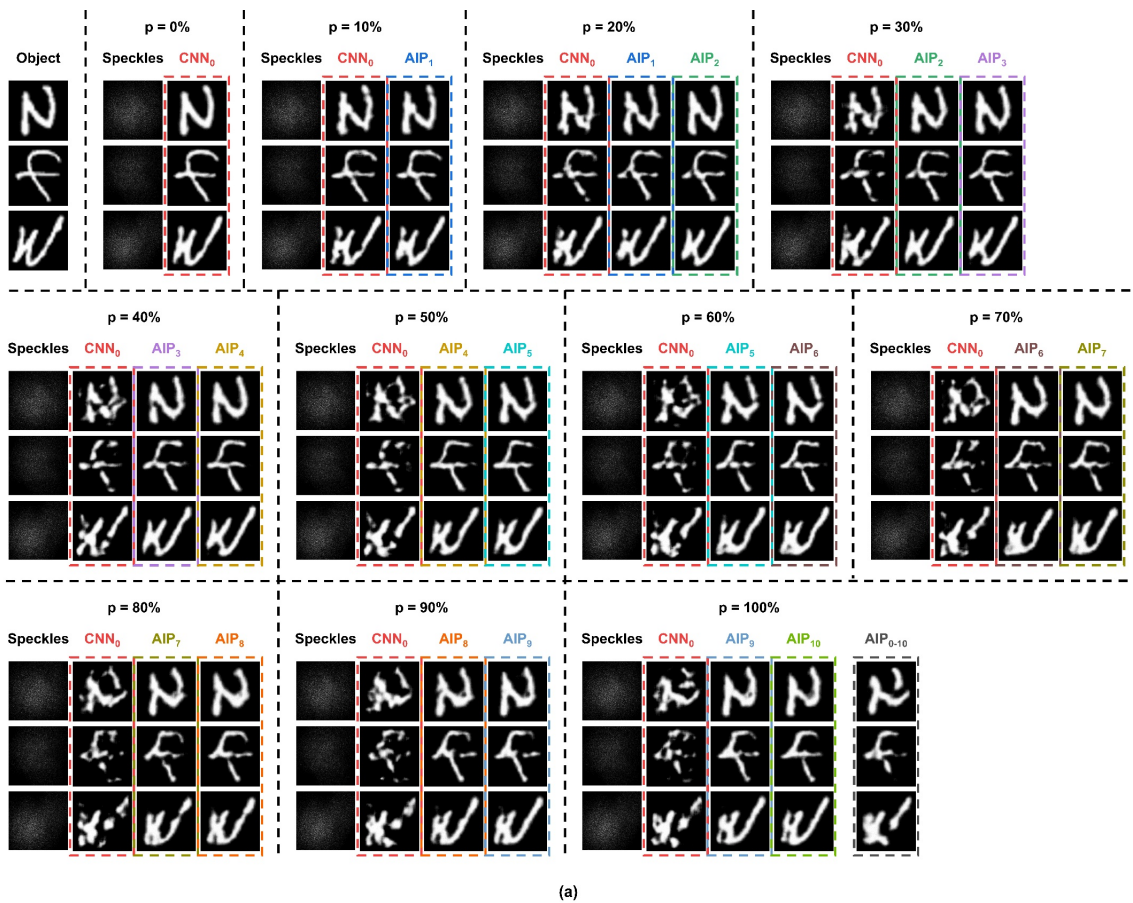

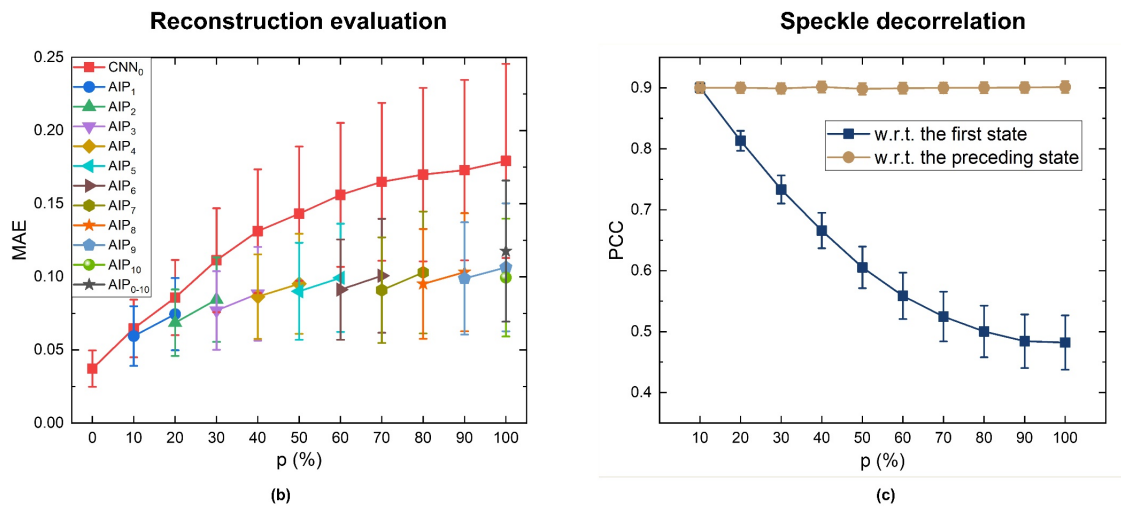

Fig. 4 (a) Image reconstructions results obtained by the AIP method to the dynamic telescopic imaging system shown in Fig.3. Object: the input object images (top left column); Speckles: the output speckle images when the percentage *p* of the substituted elements in $W(x,y)$ is increased from 0% to 100% with a step of 10%; $CNN_0$ and $AIP_i$: the reconstructed images from $CNN_0$ and the $i^{th}$ AIP. The dashed bounding box groups the reconstructions from a particular AIP or $CNN_0$. (b) The averages and standard deviations of the MAE of the test reconstructions from the $AIP_i$ and $CNN_0$. The colors of the symbols correspond to the colors of the bounding boxes in (a). (c) The PCCs of the speckle images with respect to (w.r.t.) the speckle images from the first state p=0% (dark blue line), and w.r.t. the speckle images from the preceding state (brown line).

**Dynamic GALOF-based imaging system** With the great results in the simulations shown above, we experimentally apply the AIP method to a GALOF-based imaging system[47–50]. The schematic of the setup is shown in Fig. 5 (a). We illuminate a human red blood cell sample with an LED centered at 460 nm. The cell image is first magnified by a 10x microscope objective MO1 (NA = 0.3) and a tube lens L1 (f = 200 mm), The magnified image is then sent into two arms through a beam splitter. In the reference arm, the cell image is further magnified by a combination of a 20x microscope objective MO2 (NA = 0.75) and a tube lens L2 (f = 200 mm) before being collected by a camera CDD1 (Manta G-145B, 30 fps). In the measurement arm, the cell image is passed through an 80-cm long segment of GALOF. The GALOF has a disordered structure with a diameter of 278 µm and an air-hole-filling fraction of ~28.5% (Methods). The fiber output image is magnified by the same combination of a 20x MO3 and L3 (f = 200 mm), and projected onto CCD2. By scanning the cell sample both horizontally and vertically with steps of 5 µm, we obtain numerous pairs of fiber output images and cell images. The dynamic changes of the system come from two sources. First, the cell images are moving away from the GALOF input facet (Fig. 5 (b)). At the same time, mechanical instability in the measurement arm causes small displacements among the measured fiber output images (Fig. 5 (c)). The arrows in the enlarged part of Fig. 5 (c) show the drift of the measured images during the experiment. At 0 mm imaging depth, we initialize an inverse mapping by training a $CNN_0$ on 16,000 pairs of fiber output images and cell images. After initialization, the AIP method with $m$=16,000 and $n$=1,000 is applied to stabilize the quality of reconstructed images every time when the imaging depth has increased by 1 mm. The corresponding ground truth cell images are also collected at each imaging depth for performance evaluation.

Fig. 6 (a) shows the object images, the GALOF output images, and the reconstructed images of AIPs and $CNN_0$ at each imaging depth $d$. It can be seen that $CNN_0$ generates scrambled images when dynamic changes occur. In contrast, the AIP method is able to maintain good visual qualities of reconstructed images. The differences in brightness between the reconstructed images and the objects at each imaging depth are due to the changed condition under which the object images are collected. The necessity of closely tracing the state of the system is demonstrated by $AIP_{0-3}$, where we skip the state 1 and 2, and apply the AIP method directly at the imaging depth of 3 mm. $AIP_{0-3}$ produces similar outputs all the time, i.e., mode collapse, and fails to restore images. To provide a quantitative evaluation of the AIP method regardless of the differences in brightness between the reconstructed images and the objects, we perform image segmentations (Methods) and calculate the mean intersection-over-union (IoU) scores of the 'cell' and 'background' areas (Fig. 6 (b)). The mean IoU scores of the reconstructed images are listed in Fig. 6 (a) for reference. Fig. 6 (c) shows the average values and standard deviations of the mean IoU on the test images. As can be seen, the AIP method slows down the decrease of the mean IoU scores when the imaging depth increases. Good image reconstructions with mean IoU scores larger than 0.5 are achieved up to an imaging depth of 3 mm. This agrees well with our previous work[49], where large reconstruction degradations can be observed beyond 3 mm. This can be attributed to the loss of high frequency features with increased imaging depths. However, in the previous work, we train a CNN using supervised learning at each imaging depth. The AIP method eases the burden of collecting paired images at different imaging depths while achieves comparable performance on image reconstruction.

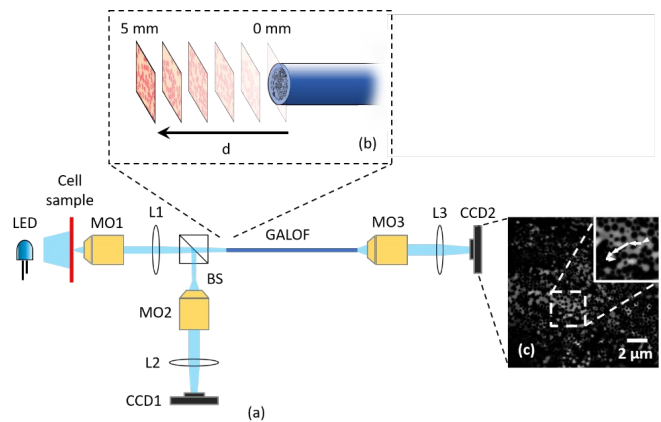

Fig. 5 (a): Schematic of the GALOF-based imaging system. MO1, MO2, MO3: microscope objectives. L1, L2, L3: tube lenses. BS: beam splitter. The imaging system is perturbed in two ways simultaneously: (b): the imaging depth from the cell image to the GALOF input facet is increased from 0 mm to 5 mm, and (c): the collected GALOF output images are drifting due to the instability in the measurement arm. The enlarged part in (c) shows the amount of drift in the collected images at the imaging depth of 0 mm to 5 mm with a step of 1 mm (the total drift is about 1 μm).

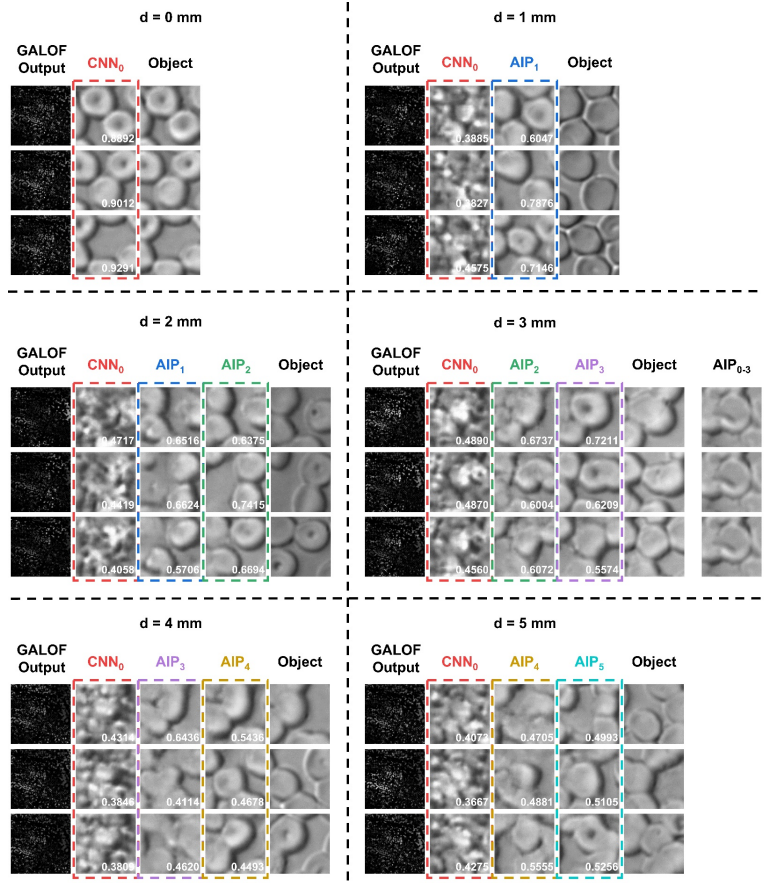

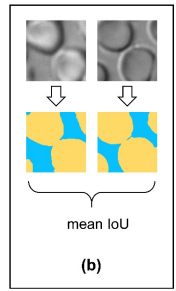

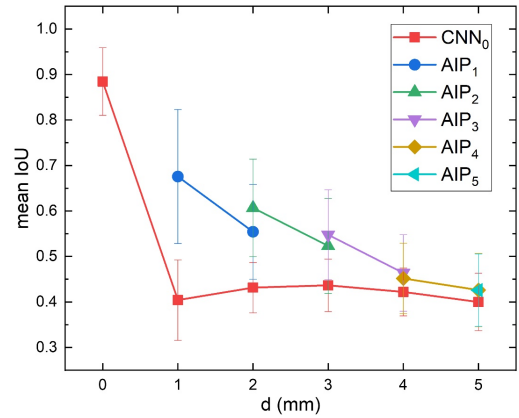

Fig. 6 (a) Image reconstructions results obtained by the AIP method to the GALOF imaging system shown in Fig.5. GALOF output: fiber output images at imaging depth $d$=0 mm to $d$=5 mm with a step of 1 mm; $CNN_0$ and $AIP_i$: reconstructed images from the $CNN_0$ and $i^{th}$ AIP; Object: human red blood cell images. The dashed bounding box groups the reconstructions from a particular AIP or $CNN_0$. In addition, a column of $AIP_{0-3}$ is shown at imaging depth $d$=3 mm to indicate the necessity of closely tracing the state of the system. The mean IoU scores of the reconstructed images are listed at the right bottom corners of the images. (b) Schematic illustration of the calculation of the mean IoU score. Here, we use the $AIP_1$ reconstructed image and the object image from the second row at $d$=1 mm as an example. The reconstructed image and object image are labeled into 'cell' (yellow) and 'background' (blue) areas. The IoU scores of the cell mask and background mask are averaged to give the mean IoU score. (c) Averages and standard deviations of the mean IoU scores calculated from the reconstructions of AIPs and $CNN_0$ images. The colors of the symbols correspond to the colors of the bounding boxes in (a).

## Discussion

**Semi-supervised learning** In the AIP method, object images are only collected at the initialization. They are used in two aspects. First, paired object images and speckle images form a training set to establish an initial inverse mapping through supervised learning. Second, a small part of object images is broadcast to the later states to stabilize the imaging performance. After the initialization, the AIP method only requires output speckle images from the dynamic scattering medium. Unsupervised learning is utilized to find a mapping from distorted reconstructed images to object images. Therefore, the AIP method eliminates the need for access to both ends of the dynamic scattering medium, as required in methods using feedback control[19–21]. This makes the AIP method easy to implement in most real-world applications, where only the output end is accessible during image acquisition.

**Flexibility** Unlike other data-driven approaches[26–30], the AIP method assumes no prior knowledge on potential dynamic changes. It adapts to the dynamic scattering medium by following the medium closely. At first glance, this remarkable performance may seem counter-intuitive, especially for the distorted images at 1 mm imaging depth for the GALOF-based imaging system (Fig. 6). Nevertheless, there is an implicit connection between the distorted images and the objects. At any state *i*, the scattering imaging system takes object images $x$ and generates speckles through a forward mapping $F_i(\cdot)$. If we take AIP$_i$ as an approximation of the system inverse mapping $F_i^{-1}$, the distorted reconstructions at state *i+1* can be represented by $F_i^{-1}(F_{i+1}(x))$. Under the condition of $F_{i+1}(\cdot) \approx F_i(\cdot)$, the translation between $x$ and $F_i^{-1}(F_{i+1}(x))$ is a 'natural' mapping for the Restore-CycleGAN to learn. Thus, the inverse mapping can be corrected if the dynamic scattering medium is traced closely.

**Universality** For linear propagation scattering media, the forward mapping operator $F_i(\cdot)$ reduces to a TM $T_i$. Thus, the dynamic changes to the medium result in different transformations of the TMs. In the first case of an evolving TM, dynamic changes simply replace elements in the TM. In the dynamic telescopic imaging system, the transformation of the TM is more implicit. In the dynamic GALOF-based imaging system, the TM $T_{i+1}$ at state *i+1* relates to the TM $T_i$ at the previous state *i* through:

$$T_{i+1} = D_{i+1} T_i P_{\Delta d_{i+1}}, \tag{3}$$

where $P_{\Delta d_{i+1}}$ is the propagation matrix, describing the free propagation of increased imaging depth $\Delta d_{i+1}$. $D_{i+1}$ is the image translation introduced by small displacements in the measurement arm. The AIP method manages to correct the inverse mappings under the TM transformations of all three cases. This shows the universality of the AIP method. We expect future successful applications of the AIP method in other scattering imaging systems.

**Perspectives** While the AIP method can correlate the inverse mapping to the dynamic scattering system, it only applies to slowly evolving systems. The speed of the current AIP method is limited by the acquisition time of collecting *m* speckle images ($10^3$ to $10^4$ depending on the complexity of the objects) to correct the inverse mapping. During the image acquisition, the AIP method assumes a quasi-static scattering system. This means that the speckle image acquisition time should be much smaller than the speckle decorrelation time. The image acquisition time is determined by two factors: the frame rate of the camera and the number of images required. While the former is limited by the hardware, much effort can be made to reduce the latter. In the current method, the image reconstruction CNN and the Restore-CycleGAN operate separately. The CNN generates distorted reconstructions and the Restore-CycleGAN finds the connection between distorted reconstructions and object images. To make more efficient use of a reduced number of speckle images, in future studies, the training of the reconstruction

CNN and the Restore-CycleGAN can be done interactively so that improvements on one lead to improvements on the other.

In conclusion, we show that the inverse mapping of dynamic scattering imaging media can be corrected through unsupervised learning if the medium is traced closely. We demonstrate the preservation of good image reconstruction by the AIP method in three showcase dynamic scattering systems. The advantages of semi-supervised learning and its flexibility make the AIP method a promising candidate to improve imaging through dynamic scattering media without prior knowledge of dynamic changes.

**Methods**

**Architectures and training processes of the Restore-CycleGAN** We use PatchGAN[38] as the discriminator network. The PatchGAN looks into patches of an input image and predicts whether they come from a real or a fake image. It consists of two input and output layers as well as five blocks in between (Extended Data Fig. 1). The last four blocks consist of Convolutional/Instance-Normalization[51]/Leaky-ReLU layers, whereas the first block omits the Instance-Normalization layer. All convolutional filters in these five blocks, except the last one, have a size of 4x4 and a stride of 2. A convolutional layer is added after these five blocks to generate the final output.

We use UNet[35] as the generator network. The UNet has an encoder-decoder architecture with skip connections between layers in the encoder and decoder (Extended Data Fig. 2). The input image is first down-sampled to a bottleneck layer by the encoder, which consists of convolutional layers with a kernel size of 4x4 and a stride of 2. The decoder then up-samples to the output image using transpose convolutional layers. Dropout layers are added to the decoder. The size of the input images is 416x416 in the GALOF-based endoscope and 256x256 in the other two systems.

All the weights in the PatchGAN and UNet are initialized through a random Gaussian distribution with a zero mean and a standard deviation of 0.02. We construct a pool of 50 fake images from the generators to train the discriminators. The pool is randomly updated through newly generated fake images. The loss function of the generator networks has four components: adversarial loss of the discriminator $\mathcal{L}_{\text{LSGAN}}$, identity loss $\mathcal{L}_{\text{identity}}$ and cycle-consistent losses $\mathcal{L}_{\text{cycle}}$ in both directions:

$$\begin{aligned}
G_{obj}^{*}, G_{corr}^{*} = \arg\min_{G_{obj}, G_{corr}} \max_{D_{obj}, D_{corr}} & \left. \mathbb{E}_{x}\left[\left(D_{obj}(x)-1\right)^{2}\right] \right. \\
& \left. + \mathbb{E}_{y}\left[D_{obj}\left(G_{obj}\left(CNN_{i}(y)\right)\right)^{2}\right] \right\} \mathcal{L}_{\text{LSGAN}}\left(G_{obj}, D_{obj}, X, Y\right) \\
& \left. + \mathbb{E}_{y}\left[\left(D_{corr}\left(CNN_{i}(y)\right)-1\right)^{2}\right] \right. \\
& \left. + \mathbb{E}_{x}\left[\left(D_{corr}\left(G_{corr}(x)\right)\right)^{2}\right] \right\} \mathcal{L}_{\text{LSGAN}}\left(G_{corr}, D_{corr}, Y, X\right) \\
& \left. + \alpha_{1}\mathbb{E}_{x}\left[\left\|G_{obj}\left(G_{corr}(x)\right)-x\right\|_{1}\right] \right. \\
& \left. + \alpha_{1}\mathbb{E}_{y}\left[\left\|G_{corr}\left(G_{obj}\left(CNN_{i}(y)\right)\right)-CNN_{i}(y)\right\|_{1}\right] \right\} \mathcal{L}_{\text{cycle}}\left(G_{obj}, G_{corr}\right) \\
& \left. + \alpha_{2}\mathbb{E}_{x}\left[\left\|G_{obj}(x)-x\right\|_{1}\right] \right. \\
& \left. + \alpha_{2}\mathbb{E}_{y}\left[\left\|G_{corr}\left(CNN_{i}(y)\right)-CNN_{i}(y)\right\|_{1}\right] \right\} \mathcal{L}_{\text{identity}}\left(G_{obj}, G_{corr}\right)
\end{aligned} \quad (4)$$

$x$ and $y$ are object and speckle images, respectively. The real images from the target domain are labeled as '1', whereas the fake images from the generator are labeled as '0'. $\alpha_1$ and $\alpha_2$ control the weighting among the losses. The four parts in the generator loss function are weighted as 1, 5, 10 and 10. We use the mean square error (MSE) for the discriminator output and the mean absolute error (MAE) for the generator output. The loss of the discriminators is weighted as one half to the loss of the generators. The discriminators and generators are trained using a batch size of 1 and an Adam optimizer with a learning rate of 0.0002 and $\beta_1$=0.5.

**Comparison between CycleGAN and Restore-CycleGAN** In Extended Data Fig. 3, we compare the restored images by CycleGAN and Restore-CycleGAN. The distorted images and object images are from the GALOF-based endoscope at an imaging depth of 1 mm. As can be seen, CycleGAN only imposes style transfer to the distorted images. In contrast, Restore-CycleGAN achieves remarkable image restorations.

**Cell image segmentation** To evaluate the image reconstruction in the GALOF-based imaging system, we perform the cell image segmentation using deep neural networks. First, the cell images are manually segmented into areas of 'cell' or 'background'. To reduce the work involved, we take advantage of the fact that the collected cell images have partial overlaps since the cell sample is scanned both horizontally and vertically in steps of 5 µm. Thus, we align each cell image according to its preceding image using the monomodal image registration and translation geometric transformation. The registered cell images are then stitched to form a panorama (Extended Data Fig. 4). After labeling the panorama, we disassemble the image mask into ground-truth segmentations for each individual cell images. In this way, we generate 220 paired cell images and segmentations. The 220 pairs of images are shuffled and divided into 180 training pairs, 20 validation pairs and 20 test pairs. We choose the neural network model and the hyper-parameters by training a model using the training set and monitoring its performance on the validation set. After a model is chosen, we combine the training and validation set to train the neural network model. Finally, the trained model is evaluated using the test set. The whole process is repeated at each imaging depth. Extended Data Table 1 shows the IoU scores of the image segmentation by the neural networks. For all the neural networks, we resize the images to 224x224 and adopt a UNet VGG19 architecture[52] with a binary cross-entropy loss function. All neural networks are trained from scratch using the Adam optimizer with a learning rate of 0.001. To perform image segmentation for the reconstructed images at different imaging depths (Fig. 6), we apply the neural network trained at 0 mm imaging depth since all reconstruction networks should generate images similar to the objects at 0 mm imaging depth.

**Generating output images through a TM** The intensity of an input image is first converted to an electric field matrix $E_{in}$. The matrix is flattened into a vector and multiplied by the complex-valued TM (Eq. 5).

The resulting vector is then rearranged to an output electric field matrix $E_{out}$. Finally, the output electric field matrix is converted back to the output intensity.

$$E_{out} = \begin{pmatrix} E_{out,1} \\ E_{out,2} \\ \vdots \\ E_{out,n} \end{pmatrix} = \begin{pmatrix} t_{11} & \cdots & t_{1n} \\ \vdots & \ddots & \vdots \\ t_{n1} & \cdots & t_{nn} \end{pmatrix} \begin{pmatrix} E_{in,1} \\ E_{in,2} \\ \vdots \\ E_{in,n} \end{pmatrix} = TE_{in} \quad (5)$$

**AIP method with *Δp*=25% on imaging through an evolving TM** We also apply the AIP method on the system shown in Fig. 2 with *Δp* increased to 25%. The results are shown in Extended Data Fig. 5. Degradation in image reconstructions can be observed when compared to the results obtained under *Δp*=12.5% (Fig. 2). This indicates the importance of tracing the state of the system closely.

**GALOF** Extended Data Fig. 6 shows a scanning electron microscope (SEM) image of the GALOF facet.

**Metrics** We use various metrics to evaluate the similarity between two images. The pixel values of these two images can be flattened to two vectors: $x$ and $y$, both containing n elements. The definitions of the metrics used in the paper are the following:

$$\text{MAE} = \frac{1}{n} \sum_{i=1}^{n} |x_i - y_i| \quad (6)$$

$$\text{MSE} = \frac{1}{n} \sum_{i=1}^{n} (x_i - y_i)^2 \quad (7)$$

$$\text{PCC} = \frac{\sum_{i=1}^{n}(x_i - \bar{x})(y_i - \bar{y})}{\sqrt{\sum_{i=1}^{n}(x_i - \bar{x})^2} \sqrt{\sum_{i=1}^{n}(y_i - \bar{y})^2}}, \quad (8)$$

where $\bar{x}$ and $\bar{y}$ are the average values of elements in $x$ and $y$, respectively. We use the mean IoU score to evaluate two image segmentations. The segmentation divides an image into masks. Each mask labels the area of one class of objects. The IoU score for two masks is

$$\text{IoU} = \frac{\text{Area of Intersection}}{\text{Area of Union}} \tag{9}$$

The mean IoU score of two image segmentations is the average IoU score over all classes.

**Data availability**

The datasets generated during the current study are available from the authors under reasonable request.

**Code availability**

The codes developed for this work are available from the authors under reasonable request.

**Extended Data**

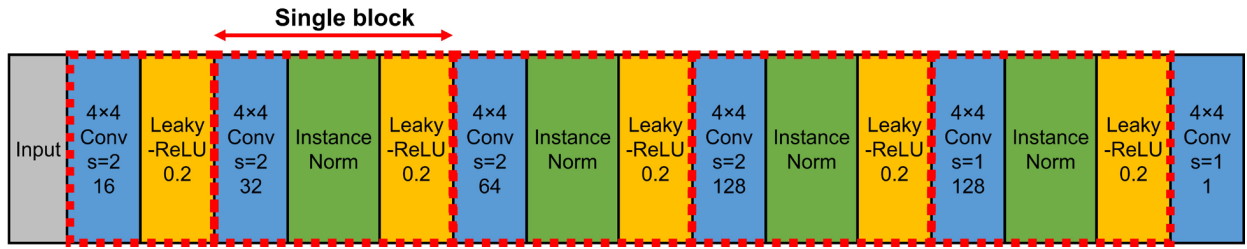

**Extended Data** Fig. 1 Architecture of the PatchGAN. Gray box: input image. Blue boxes: convolutional layers. For example, the first blue box means the convolutional layer contains 16 filters with a size of 4x4 and a stride of 2. Yellow boxes: leaky ReLU layers with a slope of 0.2. Green boxes: Instance normalization layers.

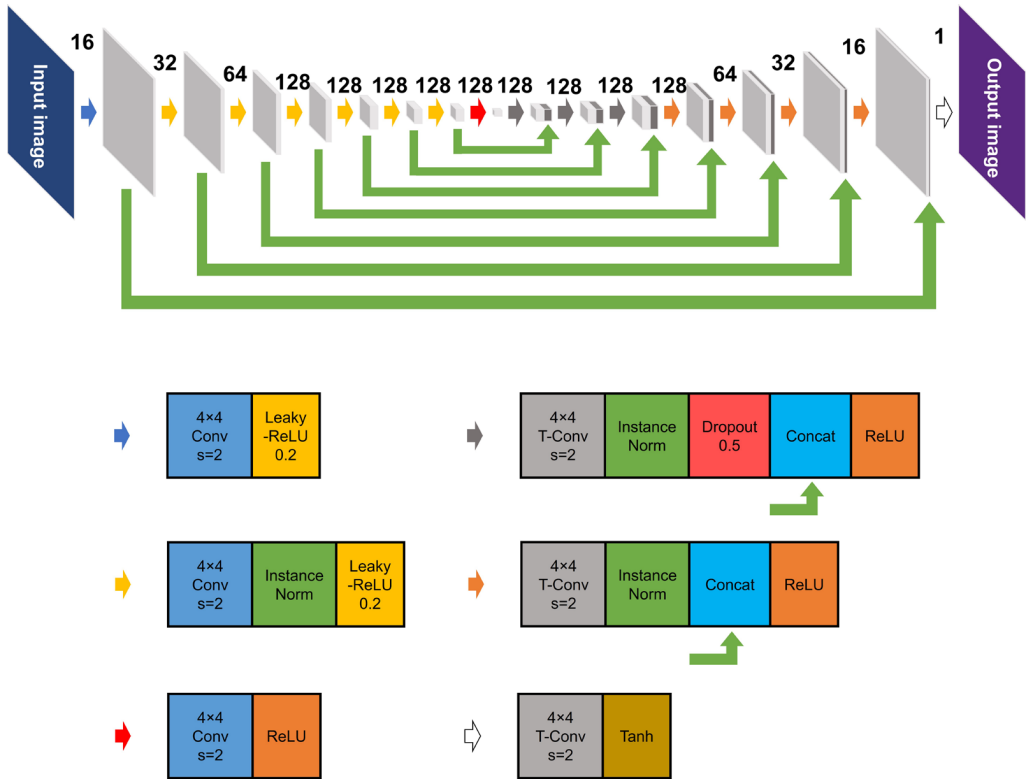

**Extended Data** Fig. 2 Architecture of the UNet. Conv: convolutional layer. T-Conv: transpose convolutional layer. Concat: concatenate operation. The numbers of filters in the convolutional layers are 16-32-64-128-128-128-128-128. The numbers of filters in the transpose convolutional layers are 128-128-128-128-64-32-16-1.

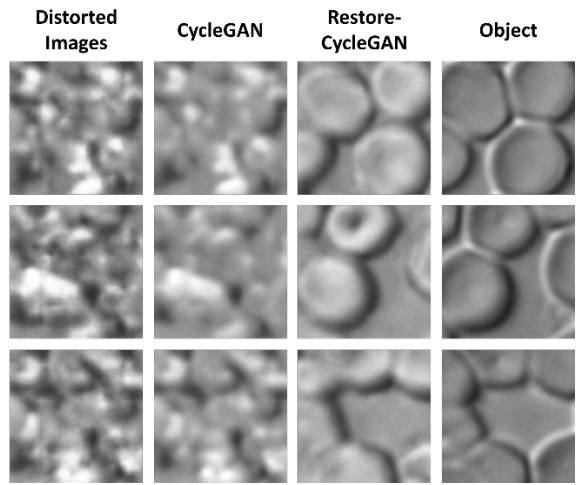

**Extended Data** Fig. 3 Comparison between CycleGAN and Restore-CycleGAN with regard to restoring distorted images. The distorted images and object images are from the GALOF-based imaging system at an imaging depth of 1 mm.

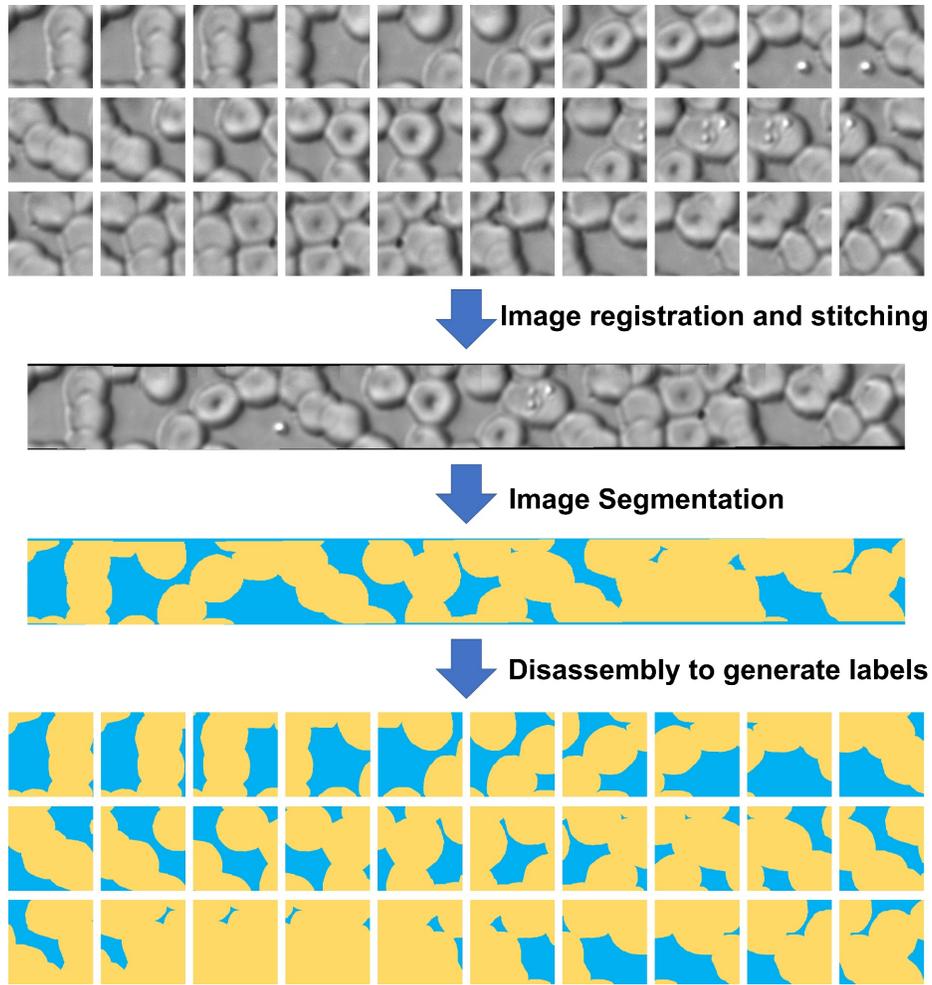

**Extended Data** Fig. 4 Flow diagram of generating ground-truth human red blood cell image segmentations. For visualization, only 30 cell images are shown here.

| Imaging depth (mm) | 0 | 1 | 2 | 3 | 4 | 5 |
|---|---|---|---|---|---|---|
| Mean IoU | 0.9457 | 0.9520 | 0.9444 | 0.8940 | 0.9621 | 0.9490 |

**Extended Data** Table 1 Mean IoU scores of the image segmentation by the neural networks at different imaging depths.

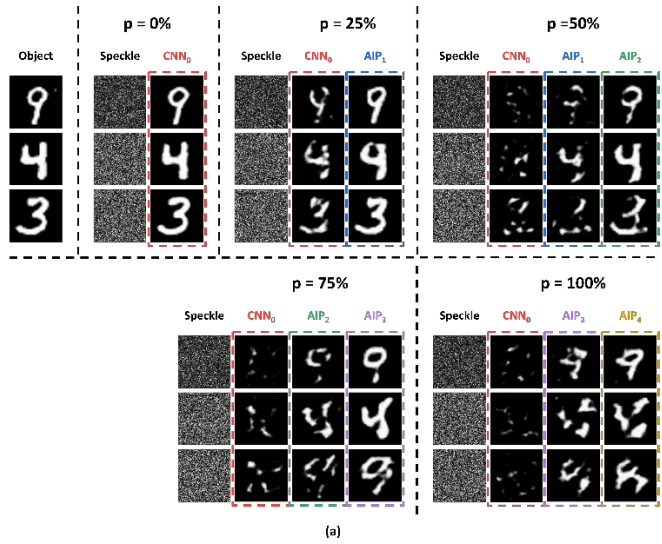

**Extended Data** Fig. 5 Results of applying the AIP method with $\Delta p$=25% to imaging through dynamic scattering shown in Fig. 2 (a). a) Object: The input object images (top left column) to the dynamic scattering system; Speckle: the output speckle images when the percentage of the substituted elements $p$ in the TM is increased from 0% to 100%; $CNN_0$ and $AIP_i$: the reconstructed images from $CNN_0$ and the $i^{th}$ AIP. (b) The averages and standard deviations of the MAE of the test reconstructions from the $AIP_i$ and $CNN_0$. The colors of the symbols correspond to the colors of the bounding boxes in (a).

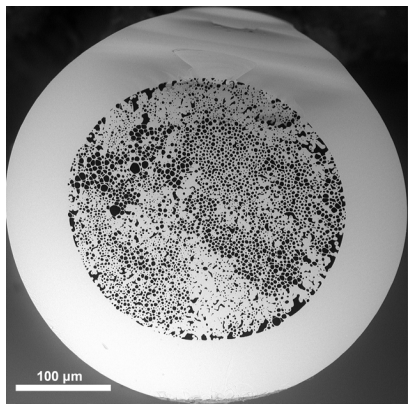

**Extended Data** Fig. 6 SEM image of the GALOF facet.

**Author Contributions**

X.H. developed the AIP framework, developed dynamic scattering imaging model, performed the numerical simulations, processed the experimental data, and wrote the first draft. J.Z. proposed the initial idea of cycle-GAN-based semi-supervised learning method for this project, supervised the project, performed the GALOF imaging experiments, and revised the manuscript. J.Z. and J.E.A. developed and fabricated the GALOF used in the experiment. S.G. assisted with the GALOF-imaging experiment. R.A.C. and A.S. supervised the GALOF development and fabrications. A.S. led the team, supervised the project, and revised the manuscript. All authors contributed to the final version of the manuscript.

**Competing interests**

The authors declare no competing interests.

**References**

1. *Wave Propagation and Scattering in Random Media*. (Elsevier, 1978). doi:10.1016/C2013-0-10906-3.
2. *Waves and Imaging through Complex Media*. (Springer Netherlands, 2001). doi:10.1007/978-94-010-0975-1.
3. Tatarskiĭ, V. I. & Silverman, R. A. *Wave propagation in a turbulent medium*. (Dover Publications, Inc, 2016).
4. Sheinin, M. & Schechner, Y. Y. The Next Best Underwater View. in *2016 IEEE Conference on Computer Vision and Pattern Recognition (CVPR)* 3764–3773 (IEEE, 2016). doi:10.1109/CVPR.2016.409.
5. Gibson, A. P., Hebden, J. C. & Arridge, S. R. Recent advances in diffuse optical imaging. *Phys. Med. Biol.* **50**, R1–R43 (2005).
6. Lee, J. S., Jurkevich, L., Dewaele, P., Wambacq, P. & Oosterlinck, A. Speckle filtering of synthetic aperture radar images: A review. *Remote Sensing Reviews* **8**, 313–340 (1994).

7.  Freund, I. Looking through walls and around corners. *Physica A: Statistical Mechanics and its Applications* **168**, 49–65 (1990).

8.  Katz, O., Small, E. & Silberberg, Y. Looking around corners and through thin turbid layers in real time with scattered incoherent light. *Nature Photon* **6**, 549–553 (2012).

9.  Goodman, J. W. *Speckle phenomena in optics: theory and applications*. (The International Society for Optical Engineering, 2020).

10. Jang, M. *et al.* Relation between speckle decorrelation and optical phase conjugation (OPC)-based turbidity suppression through dynamic scattering media: a study on in vivo mouse skin. *Biomed. Opt. Express* **6**, 72 (2015).

11. Liu, Y. *et al.* Optical focusing deep inside dynamic scattering media with near-infrared time-reversed ultrasonically encoded (TRUE) light. *Nat Commun* **6**, 5904 (2015).

12. Qureshi, M. M. *et al.* In vivo study of optical speckle decorrelation time across depths in the mouse brain. *Biomed. Opt. Express* **8**, 4855 (2017).

13. Yaqoob, Z., Psaltis, D., Feld, M. S. & Yang, C. Optical phase conjugation for turbidity suppression in biological samples. *Nature Photon* **2**, 110–115 (2008).

14. Hillman, T. R. *et al.* Digital optical phase conjugation for delivering two-dimensional images through turbid media. *Sci Rep* **3**, 1909 (2013).

15. Mosk, A. P., Lagendijk, A., Lerosey, G. & Fink, M. Controlling waves in space and time for imaging and focusing in complex media. *Nature Photon* **6**, 283–292 (2012).

16. Popoff, S. M. *et al.* Measuring the Transmission Matrix in Optics: An Approach to the Study and Control of Light Propagation in Disordered Media. *Phys. Rev. Lett.* **104**, 100601 (2010).

17. Popoff, S., Lerosey, G., Fink, M., Boccara, A. C. & Gigan, S. Image transmission through an opaque material. *Nat Commun* **1**, 81 (2010).

18. Kim, M., Choi, W., Choi, Y., Yoon, C. & Choi, W. Transmission matrix of a scattering medium and its applications in biophotonics. *Opt. Express* **23**, 12648 (2015).

19. Tajahuerce, E. *et al.* Image transmission through dynamic scattering media by single-pixel photodetection. *Opt. Express* **22**, 16945 (2014).

20. Conkey, D. B., Caravaca-Aguirre, A. M. & Piestun, R. High-speed scattering medium characterization with application to focusing light through turbid media. *Opt. Express* **20**, 1733 (2012).

21. Caravaca-Aguirre, A. M., Niv, E., Conkey, D. B. & Piestun, R. Real-time resilient focusing through a bending multimode fiber. *Opt. Express* **21**, 12881 (2013).

22. LeCun, Y., Bengio, Y. & Hinton, G. Deep learning. *Nature* **521**, 436–444 (2015).

23. Horisaki, R., Takagi, R. & Tanida, J. Learning-based imaging through scattering media. *Opt. Express* **24**, 13738 (2016).

24. Sinha, A., Lee, J., Li, S. & Barbastathis, G. Lensless computational imaging through deep learning. *Optica* **4**, 1117 (2017).

25. Li, S., Deng, M., Lee, J., Sinha, A. & Barbastathis, G. Imaging through glass diffusers using densely connected convolutional networks. *Optica* **5**, 803 (2018).

26. Li, Y., Xue, Y. & Tian, L. Deep speckle correlation: a deep learning approach toward scalable imaging through scattering media. *Optica* **5**, 1181 (2018).

27. Borhani, N., Kakkava, E., Moser, C. & Psaltis, D. Learning to see through multimode fibers. *Optica* **5**, 960 (2018).

28. Sun, Y., Shi, J., Sun, L., Fan, J. & Zeng, G. Image reconstruction through dynamic scattering media based on deep learning. *Opt. Express* **27**, 16032 (2019).

29. Fan, P., Zhao, T. & Su, L. Deep learning the high variability and randomness inside multimode fibers. *Opt. Express* **27**, 20241 (2019).

30. Li, Y., Cheng, S., Xue, Y. & Tian, L. Displacement-agnostic coherent imaging through scatter with an interpretable deep neural network. *Opt. Express* **29**, 2244 (2021).

31. Feng, S., Kane, C., Lee, P. A. & Stone, A. D. Correlations and Fluctuations of Coherent Wave Transmission through Disordered Media. *Phys. Rev. Lett.* **61**, 834–837 (1988).

32. Freund, I., Rosenbluh, M. & Feng, S. Memory Effects in Propagation of Optical Waves through Disordered Media. *Phys. Rev. Lett.* **61**, 2328–2331 (1988).

33. Katz, O., Heidmann, P., Fink, M. & Gigan, S. Non-invasive single-shot imaging through scattering layers and around corners via speckle correlations. *Nature Photon* **8**, 784–790 (2014).

34. Zhu, J.-Y., Park, T., Isola, P. & Efros, A. A. Unpaired Image-to-Image Translation using Cycle-Consistent Adversarial Networks. *arXiv:1703.10593 [cs]* (2020).

35. Ronneberger, O., Fischer, P. & Brox, T. U-Net: Convolutional Networks for Biomedical Image Segmentation. *arXiv:1505.04597 [cs]* (2015).

36. He, K., Zhang, X., Ren, S. & Sun, J. Deep Residual Learning for Image Recognition. *arXiv:1512.03385 [cs]* (2015).

37. Ulyanov, D., Vedaldi, A. & Lempitsky, V. Deep Image Prior. *Int J Comput Vis* **128**, 1867–1888 (2020).

38. Isola, P., Zhu, J.-Y., Zhou, T. & Efros, A. A. Image-to-Image Translation with Conditional Adversarial Networks. *arXiv:1611.07004 [cs]* (2018).

39. Goodman, J. W. *Statistical optics*. (John Wiley & Sons Inc, 2015).

40. Garcia, N. & Genack, A. Z. Crossover to strong intensity correlation for microwave radiation in random media. *Phys. Rev. Lett.* **63**, 1678–1681 (1989).

41. Universality classes and fluctuations in disordered systems. *Proc. R. Soc. Lond. A* **437**, 67–83 (1992).

42. Xu, J., Ruan, H., Liu, Y., Zhou, H. & Yang, C. Focusing light through scattering media by transmission matrix inversion. *Opt. Express* **25**, 27234 (2017).

43. LeCun, Y. & Cortes, C. MNIST handwritten digit database. (2010).

44. Mertz, J., Paudel, H. & Bifano, T. G. Field of view advantage of conjugate adaptive optics in microscopy applications. *Appl. Opt.* **54**, 3498 (2015).

45. Li, J. *et al.* Conjugate adaptive optics in widefield microscopy with an extended-source wavefront sensor. *Optica* **2**, 682 (2015).

46. Cohen, G., Afshar, S., Tapson, J. & van Schaik, A. EMNIST: an extension of MNIST to handwritten letters. *arXiv:1702.05373 [cs]* (2017).

47. Zhao, J. *et al.* Image Transport Through Meter-Long Randomly Disordered Silica-Air Optical Fiber. *Sci Rep* **8**, 3065 (2018).

48. Zhao, J. *et al.* Deep Learning Imaging through Fully-Flexible Glass-Air Disordered Fiber. *ACS Photonics* **5**, 3930–3935 (2018).

49. Zhao, J. *et al.* Deep-learning cell imaging through Anderson localizing optical fiber. *Adv. Photon.* **1**, 1 (2019).

50. Hu, X. *et al.* Learning-Supported Full-Color Cell Imaging Through Disordered Optical Fiber. in *Conference on Lasers and Electro-Optics* SM2L.5 (OSA, 2020). doi:10.1364/CLEO_SI.2020.SM2L.5.

51. Ulyanov, D., Vedaldi, A. & Lempitsky, V. Instance Normalization: The Missing Ingredient for Fast Stylization. *arXiv:1607.08022 [cs]* (2017).

52. Simonyan, K. & Zisserman, A. Very Deep Convolutional Networks for Large-Scale Image Recognition. *arXiv:1409.1556 [cs]* (2015).